\documentclass[sn-mathphys,Numbered]{sn-jnl}

\usepackage{graphicx}%
\usepackage{multirow}%
\usepackage{amsmath,amssymb,amsfonts}%
\usepackage{amsthm}%
\usepackage{mathrsfs}%
\usepackage[title]{appendix}%
\usepackage[dvipsnames]{xcolor}
\usepackage{textcomp}%
\usepackage{manyfoot}%
\usepackage{booktabs}%
\usepackage{algorithm}%
\usepackage{algorithmicx}%
\usepackage{algpseudocode}%
\usepackage{listings}%

\usepackage{cancel}
\usepackage{hyperref}
\usepackage{url}
\usepackage[colorinlistoftodos,textwidth=4cm,shadow]{todonotes}

\newcounter{Igor}

\newcounter{Matthias}

\begin{document}

\title[Arbitrary Controlled Re-Orientation of a Spinning Body by Evolving its Tensor of Inertia]{Arbitrary Controlled Re-Orientation of a Spinning Body by Evolving its Tensor of Inertia}


\author*[1]{\fnm{Igor} \sur{Ostanin}}\email{i.ostanin@utwente.nl}


\author[2]{\fnm{Matthias} \sur{Sperl}}\email{matthias.sperl@dlr.de}

\affil*[1]{\orgdiv{Multi-Scale Mechanics (MSM), Faculty of Engineering Technology, MESA+}, \orgname{University of Twente}, \orgaddress{\street{P.O. Box 217}, \city{Enschede}, \postcode{7500 AE}, \country{the Netherlands}}}

\affil[2]{\orgdiv{Institute of Materials Physics in Space}, \orgname{German Aerospace Center DLR}, \orgaddress{ \city{Cologne}, \postcode{51170}, \country{Germany}}}


\abstract{Bodies with the nonspherical tensor of inertia exhibit a variety of rotational motion patterns, including chaotic motion, stable periodic (quasi-periodic) rotation, unstable rotation around the direction close to the body's second principal axis, featuring a well-known tennis-racket (also known as Garriott-Dzhanibekov \cite{Trivailo2022}) effect -- series of seemingly spontaneous 180 degrees flips. These patterns are even more complex if the body's tensor of inertia (TOI) is changing with time. Changing a body's TOI has been discussed recently as a tool to perform controllable Garriott-Dzhanibekov flips and similar maneuvers. In this work, the optimal control of the TOI of the body (spacecraft, or any other device that admits free rotation in three dimensions) is used as a means to perform desirable re-orientations of a body with respect to its angular velocity. Using the spherical TOI as the initial and final point of the maneuver, we optimize the parameters of the maneuver to achieve and stabilize the desired orientation of the body's principal axes with respect to spin angular velocity. It appears that such a procedure allows for finding arbitrarily complex maneuver trajectories of a spinning body. In particular, intermediate axis instability can be used to break the alignment of the body's principal axis and the axis of rotation. Such maneuvers do not require utilization of propellants and could be straightforwardly used for attitude control of a spin-stabilized spacecraft. The capabilities of such a method of angular maneuvering are demonstrated in numerical simulations.}




\maketitle

\section{Introduction} \label{Intro}

From antiquity, humankind possesses empirical knowledge of how to manipulate the dynamic rotational motion of the body or mechanism by adjusting its mass distribution. Some particularly impressive examples can be found in acrobatic sports – martial arts, figure skating, synchronized diving, etc. However, this large array of practical knowledge was collected in presence of gravity, complicating observation of three-dimensional rotations of solid bodies due to insufficient available observation time. In more modern times, gyroscope frames and drop towers facilitated some systematic research on the topic. The beginning of the era of orbital spaceflights sparked a new wave of interest in these phenomena, however, direct ``trial and error'' research in space still remains too expensive for the implementation of exhaustive experimental programs on the topic.

It is worth noting that the fundamental equations of rigid body dynamics, suggesting different types of rotational motion control, were discussed rather early \cite{Poinsot1834} - more than a century before the era of spaceflight, predictive numerical modeling of rigid body mechanics, and almost two decades before the emergence of the first gyroscope.

Subsequently, attitude control of a spinning body became an important challenge for aerospace technology. Up-to-date satellites, spacecraft, and other systems capable to perform major orientation maneuvering, do so by introducing external torques, using small reactive thrust engines \cite{Wertz_1978}. Such an angular positioning system can be used only a limited number of times, namely – until the propellant is fully consumed. Alternatively, the existing inertial systems (reaction wheels and similar devices) are capable to adjust or stabilize the attitude very precisely, given small drift angular velocities, but cannot be used to terminate the fast rotation of a spacecraft. These, as well as some less common systems (e.g. passive ones using the gradient of drag forces of the thin atmosphere, gravitational force gradient, yoyo de-spins, etc.), fall into two categories (see, e.g. \cite{Wertz_1978}, \cite{Hughes_2004}): 

\begin{itemize}

    \item The ones using ``external'' moments $\mathbf{M}_i^{ext}$ of different nature. The conservation of angular momentum $\mathbf{L} = \mathbf{I} \boldsymbol{ \dot{\omega}}$ in this case can be written as: \footnote{The equations (\ref{eq1}-\ref{eq3}) are sketched here for a simple illustrative case of stable rotation. An analogue of (\ref{eq3}) for an arbitrary 3D rotation is discussed below.}
    \begin{equation} \label{eq1}
        \mathbf{I} \boldsymbol{ \dot{\omega}} = \sum_i \mathbf{M}_i^{ext}
    \end{equation}
    
    \item the ones redistributing the conserving angular momentum between the main body and its special rotating mechanical parts:
    \begin{equation} \label{eq2}
    \mathbf{I} \boldsymbol{ \dot{\omega}} = - \sum_i \mathbf{I}_i \boldsymbol{ \dot{\omega}}_i 
    \end{equation}
    
\end{itemize}

Until recently, the third possibility has been largely neglected - altering magnitude and direction of the spin angular velocity by changing the TOI: 

    \begin{equation} \label{eq3}
    \mathbf{I} \boldsymbol{ \dot{\omega}} = -     \mathbf{\dot{I}} \boldsymbol{ \omega} 
    \end{equation}

The fundamental difference with (\ref{eq2}) is that the change of the tensor of inertia is achieved by symmetric redistribution of weights using forces that do not create internal moments, while the reaction wheels and other similar systems imply net \textit{rotation} of the masses by nonzero internal moments.

The changes in tensor of inertia can dramatically change the behaviour of a spinning body. The presence of the deviatoric part of the TOI causes misalignment of the angular velocity and angular momentum of a spinning body, leading to chaotic motion, except the special cases of stable periodic rotations around 1-st or 3-rd principal axes (or quasi-periodic rotation/wobbling about the axes close to these principal axes). The rotation about the direction close to the 2-nd principal axis causes so-called intermediate axis instability \cite{VanDamme2017}, leading to a well-known tennis-racket effect -- a series of quasi-periodic 180 degrees flips of the body orientation with respect to its spin angular velocity. Therefore, a controlled mass redistribution leading to transformation of 1-st or 3-rd principal axis into a 2-nd one can be an efficient tool of orientation control.  

This idea has been highlighted for the first time in \cite{Beachley1969, Beachley1971}, although the idea to use moving mass mechanisms for stabilization/de-tumbling emerged even earlier \cite{Kane1963, Edwards1974}. The method developed further, in particular, in the work \cite{Mayorova2011patent} that suggested to use Garriott-Dzhanibekov flips for controllable re-orientation of a space sail.     

This concept has been revisited in 2017 by P. Trivailo et al. \cite{Trivailo2017} who have demonstrated its feasibility in a numerical simulation. The same collective of authors have later developed and generalized it in \cite{Trivailo2022}. They indicated a few different particular maneuvers that could be accomplished by altering a body's TOI. 
However, all the existing works were dealing so far only with the particular cases of orientation control, i.e. switching between the pre-defined axes of possible stable rotation. 

In this work, we demonstrate the straightforward way to achieve \textbf{arbitrary} angular re-orientations of a spinning body with respect to its rotation axis, by changing a body's TOI. The key idea is to have a \textbf{spherical} TOI at the initial and the final moment of the maneuver. This way both initial and final state are characterized by stable periodic rotation. The  desirable angular re-orientation is then achieved by optimization of the parameters of TOI's time evolution between the initial and the final state.

In the context of spacecraft attitude control and maneuvering, such a method has a number of attractive features. In contrast with the existing systems, the system manipulating the spacecraft's TOI is capable in principle to guide the body toward an arbitrarily selected orientation with respect to its axis of rotation by redistributing the energy between a chemical storage and the kinetic energy of body's own rotation. Such angular maneuvers are achieved without spending the mass of the propellant and with zero net energy losses, other than relatively small heat losses in electrical circuits and frictional mechanical contacts. Among other important features of such method of maneuvering is the possibility to change the TOI by displacing the payload rather than dead weights, and insensitivity of the maneuver structure to the absolute value of the spacecraft's angular momentum - see the in-depth discussion below.  

In order to demonstrate the capabilities of this approach, we have developed a special simulation-guided optimization framework, based on the implementation of nonspherical particle dynamics within the open-source code MercuryDPM \cite{weinhart2020fast}. All the simulation codes presented here are freely available via MercuryDPM repository at \burl{https://www.mercurydpm.org/}.

The framework convincingly demonstrates the impressive capabilities of attitude correction by optimal control of the body's TOI. Necessary manipulations with the inertia tensor of a body do have a straightforward mechanical interpretation and can be implemented in a real spacecraft.    

The paper is organized as follows. Section \ref{Methods} provides the necessary theoretical background and discusses the methodology used; the details related to numerical methods used are presented in the supplementary material. Section \ref{Results} discusses the results of numerical simulations, demonstrating the validity of our approach. Section \ref{Discussion} gives the larger picture of the fundamental value and possible applications of our findings.
        
\section{Methods} \label{Methods}

In  this section we will consider the major components of the suggested methodology of optimal control of angular orientation of a spinning body. Finer technical details are discussed in the Supplementary information.  


\paragraph{Maneuvering by changing the TOI -- general considerations}

Unlike the total mass of an isolated mass distribution (body or mechanism), its TOI can in principle be changed by altering its geometry. Hereafter we will still use the term ``rigid'' for the motion of the body that changes its TOI, although the use of such a terminology becomes ambiguous. 

By ``change'' of the TOI here and below we will understand the change in its principal components. It is important to note, however, that simple geometric considerations show that principal components of inertia can not be changed independently. For example, scaling the mass distribution along $1^{st}$ principal direction affects both $I_2$ and $I_3$. It, therefore, makes sense to choose the control parameters as mass distribution scaling factors:

\begin{equation} \label{eq4}
\begin{split}
	I_1 &= \frac{I_0}{2} (q_2(t)^2 + q_3(t)^2), \\ 
	I_2 &= \frac{I_0}{2} (q_1(t)^2 + q_3(t)^2), \\
	I_3 &= \frac{I_0}{2} (q_1(t)^2 + q_2(t)^2),
\end{split}
\end{equation}

It is also easy to see that without loss of generality we can accept $q_3(t) = 1$, as it would only contribute to a scaling multiplier of the angular velocity (see, e.g. \cite{Richter2006}). The scaling of the principal components of the TOI and their time derivatives is then given as:

\begin{equation} \label{eq5}
\begin{split}
	I_1 &= \frac{I_0}{2} (1 + q_2(t)^2), \dot{I_1} = I_0 q_2(t)\dot{q_2}(t)\\ 
	I_2 &= \frac{I_0}{2} (1 + q_1(t)^2), \dot{I_2} = I_0 q_1(t)\dot{q_1}(t)\\
	I_3 &= \frac{I_0}{2} (q_1(t)^2 + q_2(t)^2), \dot{I_3} = I_0 (q_1(t)\dot{q_1}(t) + q_2(t)\dot{q_2}(t))\\
\end{split}
\end{equation}

\begin{figure}
    \centering
    \includegraphics[width=\textwidth]{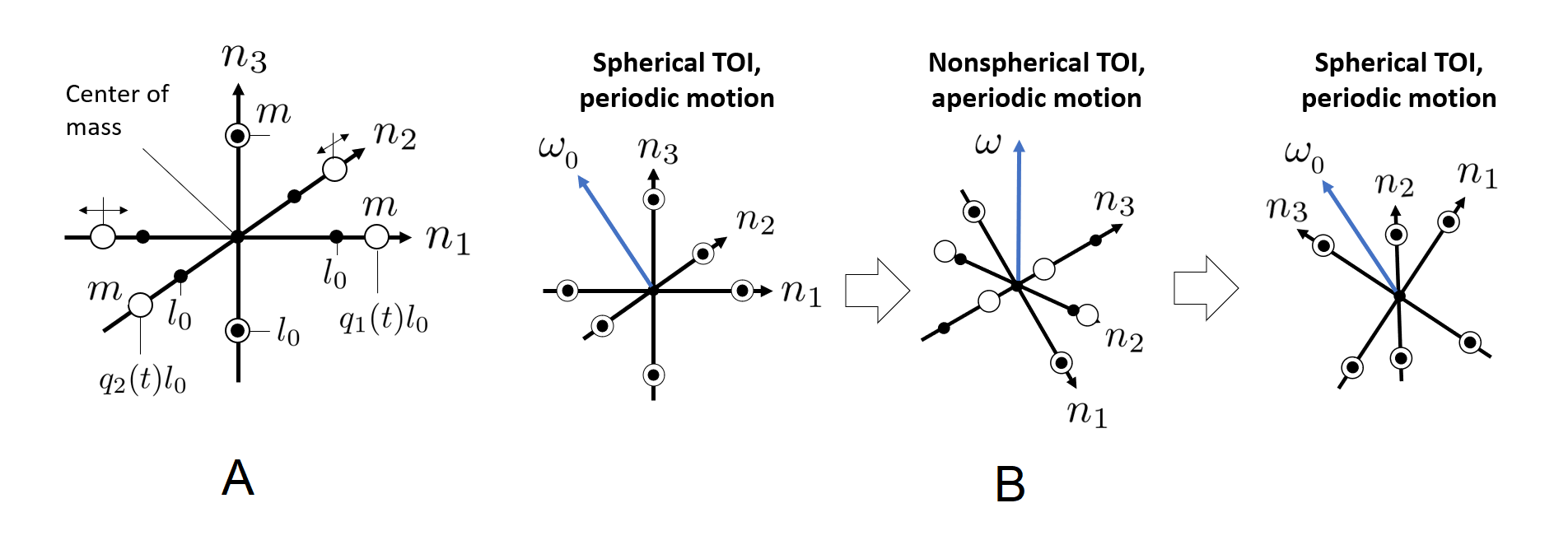}
    \caption{A) Geometric interpretation of control parameters $q_1(t), q_2(t)$. For the system on the image, $I_0 = 2ml_0^2$. (B) The schematics of an arbitrary angular maneuver.}
    \label{fig:1}
\end{figure}

Here $q_1(t)$ and $q_2(t)$ are independent control parameters that can be manipulated within a certain range between $q_{min}<1$ and $q_{max}>1$, to achieve the desirable maneuvering; $I_0$ is the baseline spherical tensor of inertia. Note that in case when $q_1(t) = 1, q_2(t) = 1$, the TOI is spherical. Fig. \ref{fig:1}(A) offers a simple mechanical interpretation of the coefficients $q_1(t)$ and $q_2(t)$, highlighting one possible way of a technical implementation (alternative ways are discussed in Section \ref{Discussion}).    

It is easy to see that certain changes in the TOI of a rotating body can be achieved with zero work of centrifugal forces. For example, the body rotating precisely around its principal axis can be arbitrarily transformed (stretched, split, etc.) along this principal axis, as long as the mass distributions around the other two axes remain the same. The other changes may be associated with positive/negative work done to move masses in the field of centrifugal forces. 

Simple physical considerations lead us to the conclusion that $q_1(t)$ and $q_2(t)$ should be twice differentiable functions with bounded 2-nd derivatives, which ensures that the transformation of the TOI can be done using finite forces/power. Below we choose the profiles $q_1(t), q_2(t)$ to be the cubic splines connecting equispaced reference values (see  Section \ref{Results}).

The usual convention in rigid body mechanics is the numbering of TOI's principal components $I_1, I_2, I_3$ in the order of their decrease. In the case of changing principal components, this convention is not useful. Further below indices $1,2,3$ do not imply order, rather, the minor, major and intermediate axes are explicitly identified if necessary. Also, in case when $q_1(t) = 1, q_2(t) = 1$, any axis of rotation is the body's principal axis. However, in our terminology, we'll use the term ``principal axis'' only for the directions that remain principal directions of the body for any values of $q_1(t), q_2(t)$.   

As mentioned above, the choice of limits for $q_1(t)$ and $q_2(t)$ ensures that the spherical TOI is available. It is therefore possible to stabilize the motion around a fixed axis by making the TOI spherical. This dictates the scheme of the maneuver, depicted in Fig. \ref{fig:1}(B). Here and below we will define the ``orientation'' $(\theta, \phi)$ as the two angles determining the direction of angular velocity $\omega$ in the own spherical coordinate system of the body (defined such that $\mathbf{n}_1$ corresponds to $(\pi/2, 0)$, $\mathbf{n}_2$ -- $(\pi/2, \pi/2,)$, $\mathbf{n}_3 = \mathbf{n}_1 \times \mathbf{n}_2$). It worth noting here that the orientation defined in this way can be interpreted as the angles of ``latitude'' and ``rotation'' of a camera, directed along $n_3$, while the ``azimuth'' of a camera is given by $\omega_0t + C$ (see Fig. 6(A) and the corresponding discussion below). The maneuver starts at a certain state with the spherical TOI $I_0$, angular velocity $\omega_0$ and the orientation $(\theta_{beg}, \phi_{beg})$. In case if the initial angular velocity is not aligned with one of the principal axes, the changes in the tensor of inertia initiate complex aperiodic motion. The sequence of changes ends with the state with a spherical tensor of inertia $I_0$ again, characterized by the angular velocity $\omega_{end}$, orientation $(\theta_{end}, \phi_{end})$. The conservation of angular momentum ensures that $\omega_{end}=\omega_0$ (during the maneuver, however, the angular velocity varies). The sequence of TOI changes is found by the optimization procedure that ensures the desired $(\theta_{end}, \phi_{end})$. The optimization technique is described below. 

It is important to note that if the body rotates precisely around one of its principal axes, the changes $q_1(t), q_2(t)$ can not perturb the periodic motion. In such a case, the principal axis, aligned with the angular velocity, can be transformed into an intermediate axis, which causes instability (see, e.g. \cite{Ashbaugh1991}), and rapid development of the misalignment. Therefore, the described system of maneuvering practically does not have the deadlock states. 

The rigorous justification of the existence and uniqueness (non-uniqueness) of the sought maneuver trajectory remains beyond the scope of this work. However, our numerical results clearly demonstrate that the optimization algorithm, given proper maneuver parameters search space, always finds the maneuver leading precisely to the desired state, even for transitions between the states with close alignment of the axis of rotation with the principal axes.

\paragraph{Rotational motion of a body that changes its TOI}

Based on the considerations above, we accept the following assumptions on the rotational motion of the body changing its TOI:

\begin{itemize}
	\item We consider $C^2$-continuous evolution of TOI's principal values, given by (\ref{eq5}). 
	
	\item The TOI and its first time derivative are prescribed precisely in the local (rotating) coordinate system at every moment of time.
\end{itemize}  

The equations of motion are obtained straightforwardly by generalization of the standard derivation of Euler's equations of rigid body rotation. These equations are obtained from the condition of conservation of angular momentum in the absence of external moments:

\begin{equation} \label{eq6}	
	\dot{ \mathbf{L}} = 0  	
\end{equation}

Expansion of the time derivative in (\ref{eq6}) leads to the following equation for the rotational motion of a body changing its TOI in the local (rotating) frame:

\begin{equation} \label{eq7}	
	\dot{\mathbf{I}}^l(t) \boldsymbol{\omega}^l(t) + \mathbf{I}^l(t) \dot{\boldsymbol{\omega}}^l(t) + \boldsymbol{\omega}^l(t) \times \mathbf{I}^l(t)\boldsymbol{\omega}^l(t) = 0   
\end{equation}

The derivation of this equation, its coordinate form in the inertial frame and the employed algorithm of its numerical solution are discussed in the Supplementary Information.

\paragraph{Dimensionless system of units used}

It is natural to introduce the dimensionless quantities characterizing the maneuver. Moments of inertia are further measured in $I_0$, and angular velocities in $\omega_0$. This naturally introduces units of time ($t_0 = 2 \pi / \omega_0$), angular momentum ( $I_0 \omega_0$ ) and energy ( $I_0 \omega_0^2/2$ ). The remaining quantities characterizing the system ($N, q_1, q_2$) are dimensionless. The dimensionless duration of the maneuver $T = (t_{end} - t_{beg})/t_0$, number of reference points $N$ and the span $[q_{min}, q_{max}]$ define the parameter space where the optimal maneuver is sought.

\paragraph{Simulation-guided optimization procedure}

The procedure to perform an optimization-based search for the control parameters providing the desired maneuver is similar to the one recently suggested by the author and his colleagues in the work \cite{Ostanin2022}. We seek to find the control parameters $q_1(t), q_2(t)$, providing the maneuver highlighted in Fig. \ref{fig:1}(B). 

In order to guide the body toward the desired final orientation $(\theta, \phi)$ , we use the following definition of the functional:

\begin{equation} \label{eq8}
\begin{split}
    \mathcal{L}(\theta, \phi, \theta_{goal}, \phi_{goal}) &= \arccos(\mathbf{p}(\theta, \phi) \mathbf{p}_{goal}(\theta_{goal}, \phi_{goal})), \\
    \mathbf{p}(\theta, \phi) &= \begin{pmatrix}
		\sin{\theta}\cos{\phi} \\
		\sin{\theta}\sin{\phi} \\
		\cos{\theta} \\
		\end{pmatrix}, \\
	\mathbf{p}_{goal}(\theta_{goal}, \phi_{goal}) &= \begin{pmatrix}
		\sin{\theta_{goal}}\cos{\phi_{goal}} \\
		\sin{\theta_{goal}}\sin{\phi_{goal}} \\
		\cos{\theta_{goal}} \\
		\end{pmatrix}. \\
\end{split}
\end{equation}
     
i.e., we seek to minimize the angle between the current and the desired orientation. Such functional definition does not penalize for the duration of the maneuver, complexity or rate of change of the tensor of inertia, therefore, these parameters should be prescribed as to ensure the feasibility of the maneuver. This definition also does not penalize for the energy one needs to ``borrow'' to accomplish the maneuver and the number of control reference points used.     
This functional is strictly zero once the body's final orientation $(\theta, \phi)$ precisely matches the goal orientation $(\theta_{goal}, \phi_{goal})$. 

The time evolution of coefficients $q_1(t),q_2(t)$ is given by the cubic splines (implemented in \cite{2020SciPy}). The initial and final nodal values $q_i(t_{beg}), q_i(t_{end})$ are fixed to $1$ (spherical tensor of inertia), initial and final time derivatives $\dot{q}_1(t), \dot{q}_2(t)$ are fixed to zero. The remaining $2N$ nodal values $q = (q_1^1..q_1^{N},q_2^{1}..q_2^{N})$ are varied in an unbounded and unconstrained multidimensional optimization procedure. The optimizer seeks for a vector of unknowns $\mathbf{X}: X_i \in \mathcal{R}$, which are mapped to $ \mathbf{q}: q_i \in [q_{min}, q_{max}]$ in the following way:

\begin{equation} \label{eq9}
	\mathbf{q}(\mathbf{X}) = \frac{q_{max}+q_{min}}{2} - \frac{q_{max}-q_{min}}{2} \cos{\mathbf{X}}
\end{equation}

The Powell optimization algorithm \cite{Powell1964}, as implemented in \cite{2020SciPy} is employed to vary the control parameters evolution. The maneuver duration $T$, the number of reference values $N$ and the range $[q_{min}, q_{max}]$ are chosen empirically outside of the optimization cycle.

\begin{figure} 
    \centering
    \includegraphics[width=\textwidth]{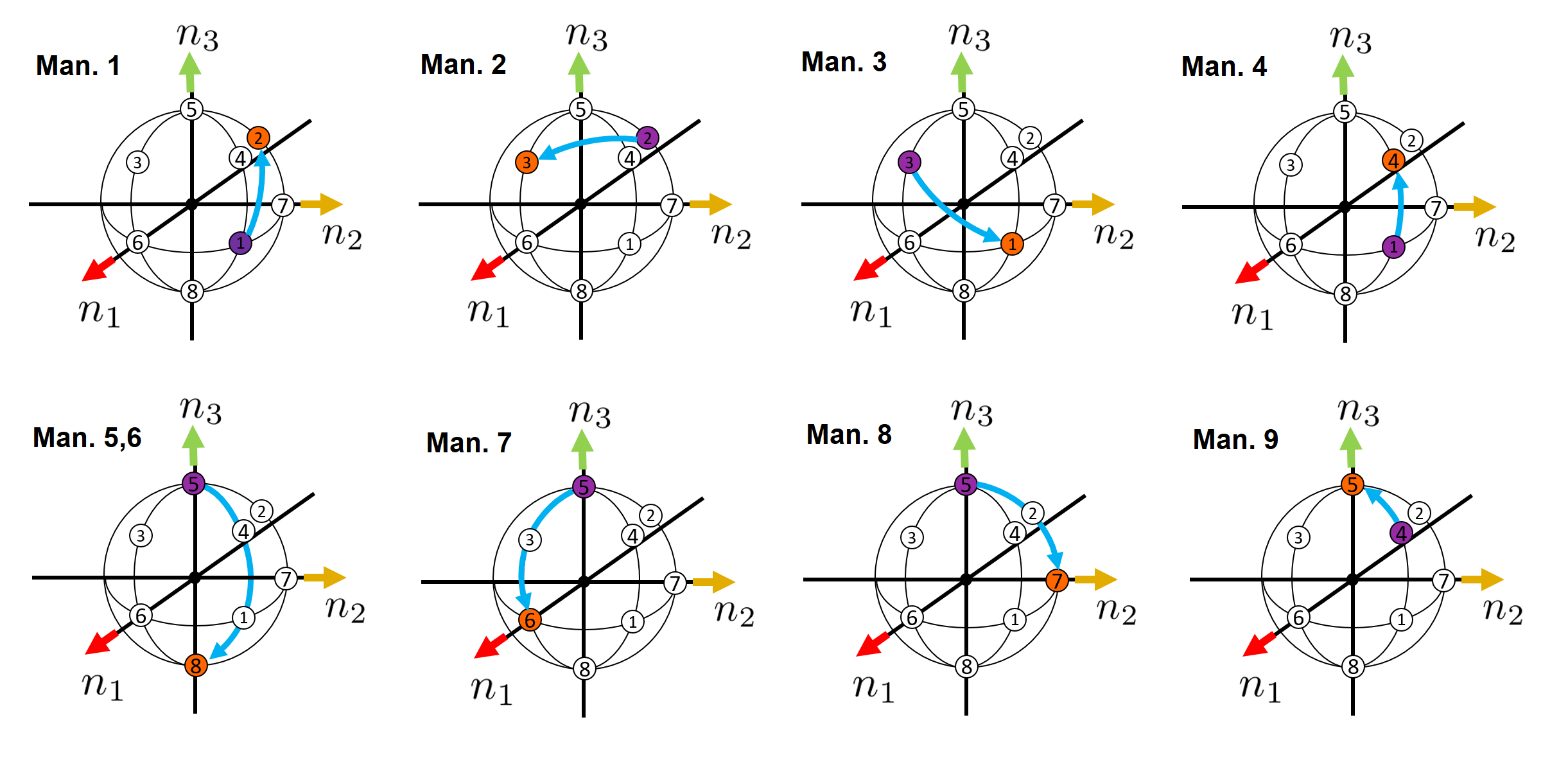}
    \caption{Schematic representations of paths of re-orientation maneuvers studied. See Table 1 for explicit specification of points coordinates, and  Table 2 for the extended description of the maneuvers studied.}
    \label{fig:2}
\end{figure}

\section{Results} \label{Results}

As a brief illustration of the suggested approach, let us consider the optimization of TOI control parameters to perform  arbitrary angular re-orientations of a spinning body.

The dimensionless system of units described above is used. Unless otherwise noted, the control parameters are varied between $0.5$ and $1.5$, leading to the ranges for principal moments of inertia:

\begin{equation} \label{eq10}
	I_1 \in (0.625,1.625), \
	I_2 \in (0.625,1.625), \
	I_3 \in (0.25, 2.25).
\end{equation}

The kinetic energy can therefore vary in the range $E \in (0.222, 2.0)$, while the angular momentum is constant and equal to $1$. \

Every maneuver is the transition between the orientation $i:(\theta_i, \phi_i)$ and orientation $j:(\theta_j, \phi_j)$. As was discussed above, the initial and final points of the maneuver are characterized by unit spherical TOI, while during the maneuver TOI varies. Initial and final orientations can be visualized as points on a unit sphere - see Fig. \ref{fig:2}. Table 1 gives the set of points that was chosen to demonstrate the capabilities of the method.

\begin{table}
\centering
	\label{tab1}
	\caption{Reference orientations, given in terms of angles $(\theta, \phi)$ and unit direction vectors $\mathbf{p}$.}
\begin{tabular}{|c|c|c|c|}
	\hline 
	Point & $\theta$ & $\phi$  & $\mathbf{p}$ \tabularnewline
	\hline
	\hline 
	1 & $\pi/2$ & $\pi/4$ & $2^{-1/2}(1,1,0)$ \tabularnewline
	\hline
	2 & $\pi/4$ & $\pi/2$ & $2^{-1/2}(0,1,1)$ \tabularnewline
	\hline
	3 & $\pi/4$ & $0$ & $2^{-1/2}(1,0,1)$ \tabularnewline
	\hline 
	4 & $\arccos{(3^{-1/2})}$ & $\pi/4$ & $3^{-1/2}(1,1,1)$ \tabularnewline
	\hline 
	5 & $0$ & $0$ & $(0,0,1)$ \tabularnewline
	\hline 
	6 & $\pi/2$ & $0$ & $(1,0,0)$ \tabularnewline
	\hline 
	7 & $\pi/2$ & $\pi/2$ & $(0,1,0)$ \tabularnewline
	\hline 
	8 & $\pi$ & $0$ & $(0,0,-1)$ \tabularnewline
	\hline 
\end{tabular}
\end{table}

Table 2 summarizes the list of maneuvers considered in this section. The first column gives maneuver number. The second column gives the maneuver path in terms of reference orientations specified in Table 1. The third column details whether the maneuver is the result of optimization(``O'') or a run with the prescribed control parameters (``P''). The fourth column gives the number of control reference points $N$ (the number of optimization parameters is $2N$). The fifth column gives the dimensionless duration of the maneuver $T$. The sixth column gives the range of control parameters $[q_{min}, q_{max}]$. The seventh column gives the value of the goal functional after convergence of the optimization procedure. The eight column lists the number of functional evaluations during the optimization procedure (the number of simulation framework runs). The last column gives the link to the video of the maneuver (if available).

Maneuvers 1-9 illustrate the transitions between orientations that are not aligned with the principal axes of inertia. 

Maneuver 10 is the single controlled Garriott-Dzhanibekov flip, performed without parameter optimization - see below. 

Maneuvers 11-14 are transitions between the orientations close to the principal axes. 

\begin{figure} 
    \centering
    \includegraphics[width=\textwidth]{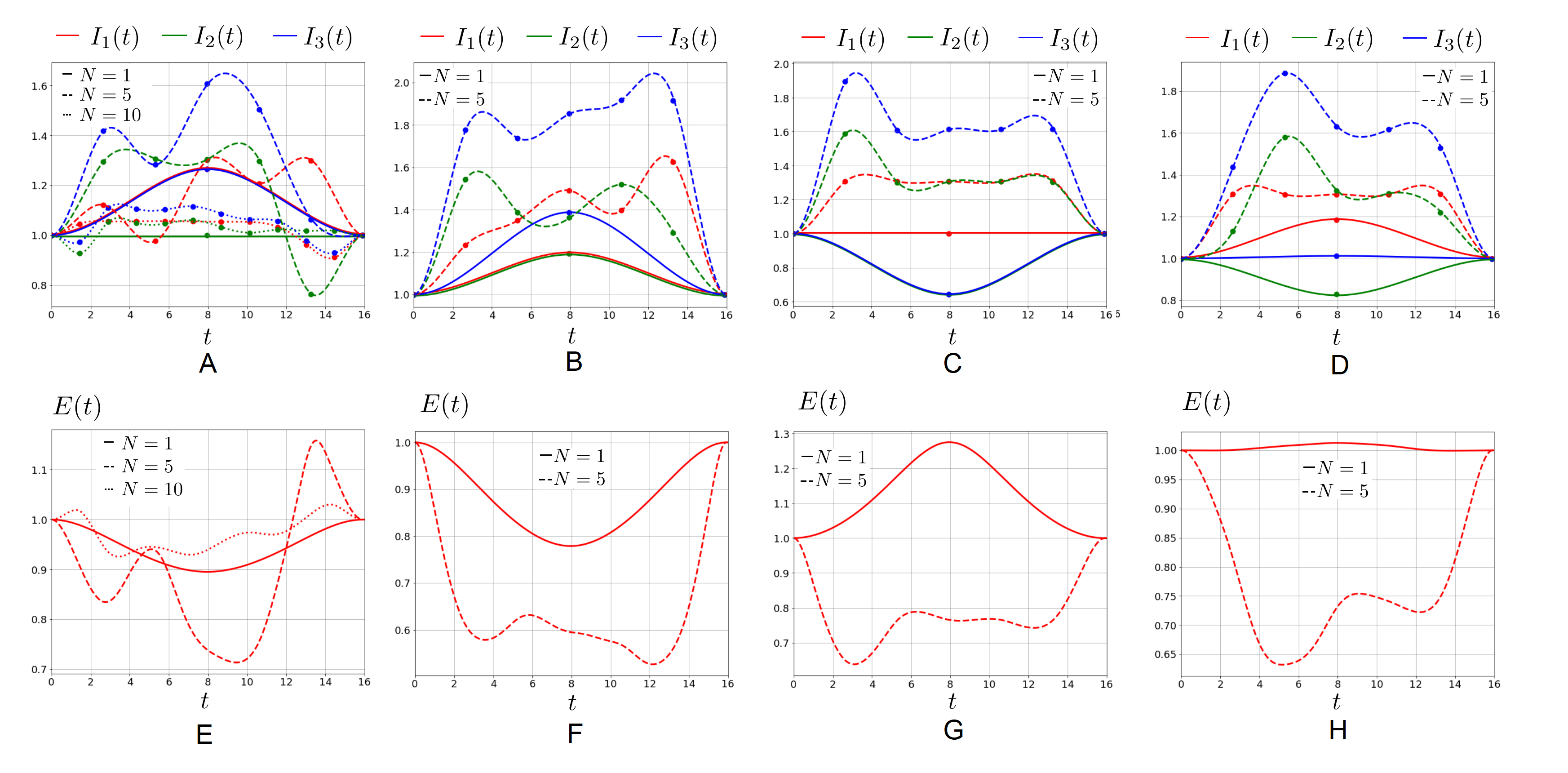}
    \caption{Time evolution of principal moments of inertia A)-D) and kinetic energies E)-H), corresponding to the A), E) -- \pmb{path 1-2} (maneuvers 1 \cite{Maneuver_1}, 5, 9) ;  B), F) -- \pmb{path 2-3} (maneuvers 2 \cite{Maneuver_2}, 6);  C), G) -- \pmb{path 3-1} (maneuvers 3 \cite{Maneuver_3}, 7) ; D), H) --
		\pmb{path 1-4} (maneuvers 4\cite{Maneuver_4}, 8)}.
    \label{fig:3}
\end{figure}

Fig. \ref{fig:3}. gives the evolution of principal moments of inertia and rotational kinetic energy during the maneuvers 1-9. All of these maneuvers were easily achievable with $T = 16$ (time corresponding to $16$ periods of rotation for $q_1=q_2=1$). We showcase the results of optimization for different numbers of reference points and range of control parameters. It appears that precise convergence of the maneuver is achievable, given sufficient time and complexity of the maneuver, as well as the range of control parameters. It appears that the reduced range of control parameters (maneuver 9) does not prevent successful maneuvering, given increased number of maneuver reference points ($N=10$). Comparison of maneuvers with the same initial and final orientations, but different parameters (e.g. different number of reference points) often indicates qualitatively different solutions -- i.e. there is a wide variety of possible maneuver trajectory leading to the desired re-orientation. Another important observation is that the kinetic energy of initial and final state matches precisely (the error does not exceed $10^{-4}$), meaning that the total work of the internal forces to accomplish the maneuver is always zero.

\begin{figure} 
    \centering
    \includegraphics[width=\textwidth]{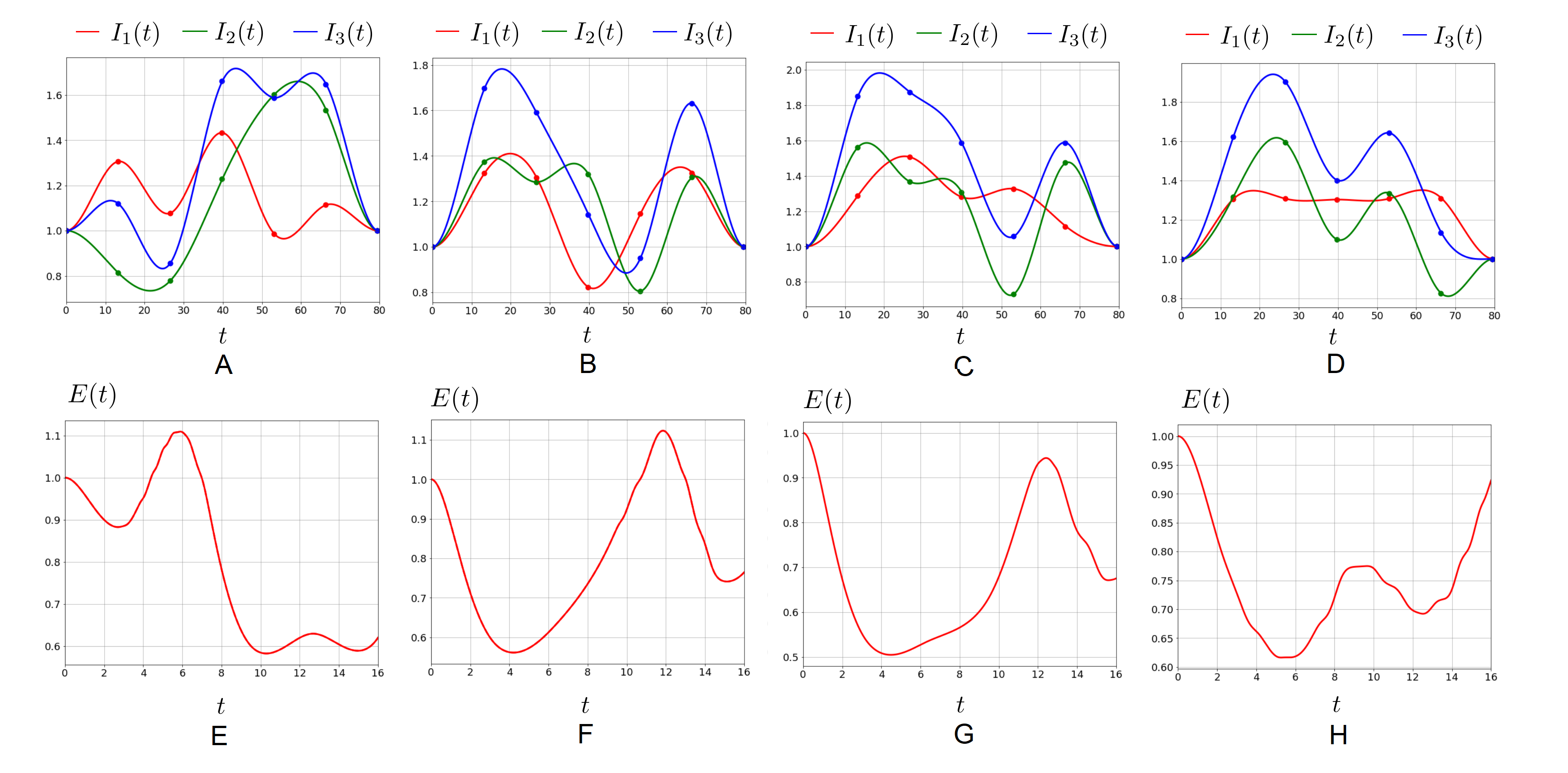}
    \caption{Time evolution of principal moments of inertia A)-D) and kinetic energies E)-H), corresponding to the A), E) -- \pmb{maneuver 11}, B), F) -- \pmb{maneuver 12},  C), G) -- \pmb{maneuver 13}, D), H) --
		\pmb{maneuver 14}.}
    \label{fig:4}
\end{figure}

Fig. \ref{fig:4} details the evolution of TOI and rotational energy for the maneuvers 11-14, whose start and/or end orientations are in close vicinity of the principal axes. For these maneuvers, the algorithm also performed beyond expectations. One can not start the maneuver if the rotation is perfectly aligned with the principal axis - any changes of $q_1,q_2$ will not induce misalignment. However, if the initial offset from the principal axis is $10^{-3}\pi$ rad, it achieves to the goal orientation with the reasonably good precision (Table 2). It worth noting that the slow development of misalignment and convergence to the orientation close to a principal axis lead to longer maneuvers -- the dimensionless duration of the maneuvers 11-14 was increased five times compared to maneuvers 1-9.

\begin{table}
\centering
	\label{tab2}
	\caption{Benchmark maneuvers and their parameters}
\begin{tabular}{|c|c|c|c|c|c|c|c|c|}
	\hline 
	Maneuver & Path & O/P  & $N$ & $T$ & $[q_{min},q_{max}]$ &$\mathcal{L}(\theta_{end}, \phi_{end})$ & $N_{fev}$ & Video \tabularnewline
	\hline
	\hline 
	1 & 1-2 & O & $1$ & $16$ & $(0.5, 1.5)$ & 0 & $434$ & \cite{Maneuver_1}  \tabularnewline
	\hline
	2 & 2-3 & O & $1$ & $16$ & $(0.5, 1.5)$& 0 & $322$ & \cite{Maneuver_2} \tabularnewline
	\hline
	3 & 3-1 & O & $1$ & $16$ & $(0.5, 1.5)$& 0 & $392$ & \cite{Maneuver_3} \tabularnewline
	\hline
	4 & 1-4 & O & $1$ & $16$ & $(0.5, 1.5)$& 0 & $771$ & \cite{Maneuver_4} \tabularnewline
	\hline
        5 & 1-2 & O & $5$ & $16$ & $(0.5, 1.5)$ & 0 & $2120$ &   \tabularnewline
	\hline
	6 & 2-3 & O & $5$ & $16$ & $(0.5, 1.5)$& 0 & $1302$ &  \tabularnewline
	\hline
	7 & 3-1 & O & $5$ & $16$ & $(0.5, 1.5)$& 0 & $1808$ &  \tabularnewline
	\hline
	8 & 1-4 & O & $5$ & $16$ & $(0.5, 1.5)$& 0 & $1280$ &  \tabularnewline
	\hline
 	9 & 1-2 & O & $10$ & $16$ & $(0.9, 1.1)$& 0 & $2487$ &  \tabularnewline
	\hline

	10 & 5-8 & P & - & $7.2$ & - & $3.340 \times 10^{-3}$ & - & \cite{Maneuver_10}\tabularnewline
	\hline
	11 & 5-8 & O & $5$ & $80$ & $(0.5, 1.5)$ & $5.227 \times 10^{-4}$  & $685$ & \tabularnewline
	\hline
	12 & 5-6 & O & $5$ & $80$ & $(0.5, 1.5)$ & $3.769 \times 10^{-2}$ & $2197$ &\tabularnewline
	\hline
 	13 & 5-7 & O & $5$ & $80$ & $(0.5, 1.5)$ & $9.661 \times 10^{-3}$ & $4312$ & \tabularnewline
	\hline
  	14 & 4-5 & O & $5$ & $80$ & $(0.5, 1.5)$ & $6.864 \times 10^{-3}$ & $2510$ & \tabularnewline
	\hline
	 
\end{tabular}
\end{table}

One interesting particular case is controlled Garriott-Dzhanibekov flip - starting from the rotation around 3-rd principal axis and spherical tensor of inertia, we transform the axis of rotation into an intermediate axis, by setting

\begin{equation} \label{eq11}
\begin{split}
q_1^2 &= a, \\
q_2^2 &= 2 - q_1^2.     
\end{split}
\end{equation}

Once the first $\pi$ rad flip is completed, we set $q_1 = 1, q_2 = 1$ again and stabilize the rotation. Both changes occur at the moment of alignment of rotation with the third principal axis, and therefore do not cost energy and do not induce misalignment. Our numerical experiments demonstrate that the idea of such maneuver is working \cite{Maneuver_10}. The weak point is that the changes in the tensor of the inertia should be instantaneous (or at least sufficiently fast)  which seems impractical for the applications. The optimization routine addressing the same task (maneuver 11) also resorts to development of tennis-racket instability, but does it in the relaxed form, with more complex program of changes of TOI. 

The optimization problem is, however, ill-posed in this case - the final set of control parameters strongly depends on small initial misalignment of the rotation axis with the body's principal axis. Same applies for all other maneuvers starting from the angular velocity close to one of the body's principal axes of inertia. Therefore, it appears to be convenient to perform such maneuvers in two steps: in the first step, certain misalignment is produced by development of tennis racket instability, resulting in the rotation ($(\theta_{c}, \phi_{c})$). At the second step, the transition $(\theta_{c}, \phi_{c}) \xrightarrow{} (\theta_{goal}, \phi_{goal})$ is achieved by solving a proper (well-posed) optimization problem. The advantage of such an approach is that the first step should not precisely define intermediate orientation $(\theta_{c}, \phi_{c})$, it is only important to achieve a certain rotation axis, significantly misaligned with the body's principal axes.

\begin{figure} 
    \centering
    \includegraphics[width=\textwidth]{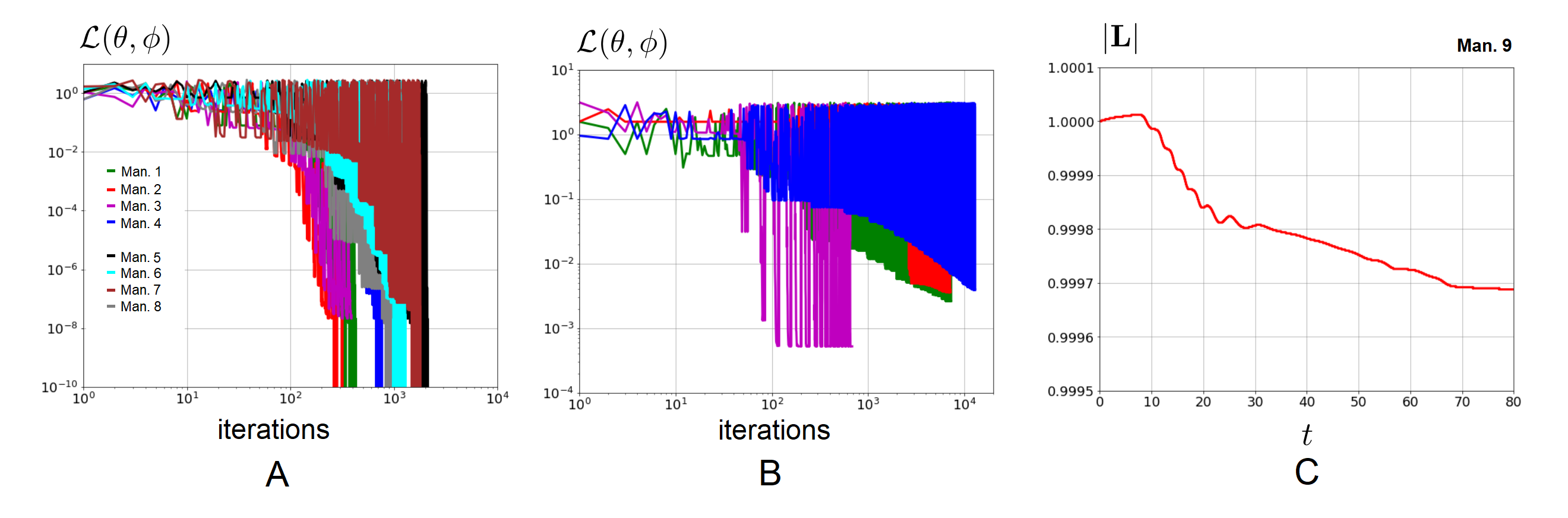}
    \caption{(A, B) Evolution of the goal functional as the function of the iteration number for (A) maneuvers 1-8, (B) maneuvers 11-14. (C) Drift of angular momentum during the simulation (Maneuver 11).}
    \label{fig:5}
\end{figure}

Our numerical experiments indicate that achieving the goal orientation co-oriented with one of the principal axes is rather challenging. Fig \ref{fig:5}(A,B) illustrates the rates of convergence of optimization algorithms for the maneuvers 1-8 and maneuvers 11-14. Note that the oscillations of the functional with iteration number is the specific feature of Powell's optimization algorithm, which finds the local minima by bidirectional search in a parameter space along certain directions that are explored sequentially. - see \cite{Powell1964} for details. 
One can clearly see that in the latter case the convergence to a local minimum is much slower, and final values are far from zero. Still, these maneuvers converged with a reasonably good precision (see Table 2).

Therefore, it is clear that the described approach to angular maneuvering is rather stable and allows to achieve any re-orientation of the spinning body with respect to its angular velocity. Initial alignment of the angular velocity of rotation with the principal axis slows down the maneuver, but can not become a deadlock, since tennis racket instability rapidly develops even from initially tiny misalignment.

Figure \ref{fig:5}(C) illustrates the quality of angular momentum conservation in our simulations. It can be seen that the quantity that should be precisely constant, if fact, features some drift during numerical motion integration -- on the order of $10^{-4}$ of its absolute value during the longest simulation time span. This can be considered as sufficiently good quality of time integration, which means trustworthiness of our simulations.

\section{Discussion and conclusions} \label{Discussion}

In our work, we approached the idea of attitude control by changing a body's tensor of inertia, that was first highlighted in \cite{Beachley1969, Beachley1971} and later developed in \cite{Trivailo2017,Trivailo2022} in the aerospace community. We have suggested a few important advances of this idea, most importantly, the idea of a maneuver that starts and ends at the state of a spherical tensor of inertia. This way any arbitrary orientation of a body with respect to its axis of rotation can be stabilized by a certain maneuver, which can be determined by the optimization procedure. This results in essentially new technology of attitude control of a spinning body. 
It is interesting to note that our orientation parameters $(\theta, \phi)$ admit a simple interpretation. The orientation of a camera or other similar payload (telescope, dipole antenna) directed outward from the center of mass, requires specifying three angular parameters in the static case. However, for a body spinning about the fixed axis, one can alter only two angular parameters, while the ``azimuth'' angle is prescribed as $\omega_{0} t + C$. Simple geometric considerations establish the identity of our orientation angles $(\theta, \phi)$ with the above-mentioned angular parameters, for the case of directed payload aligned with the axis $n_3$. Fig. 6(A) illustrates such an interpretation.

The advantage of the approach is that such maneuvers cost practically zero energy (neglecting heat losses in electric circuits and frictional contacts). 

\begin{figure} 
    \centering
    \includegraphics[width=12 cm]{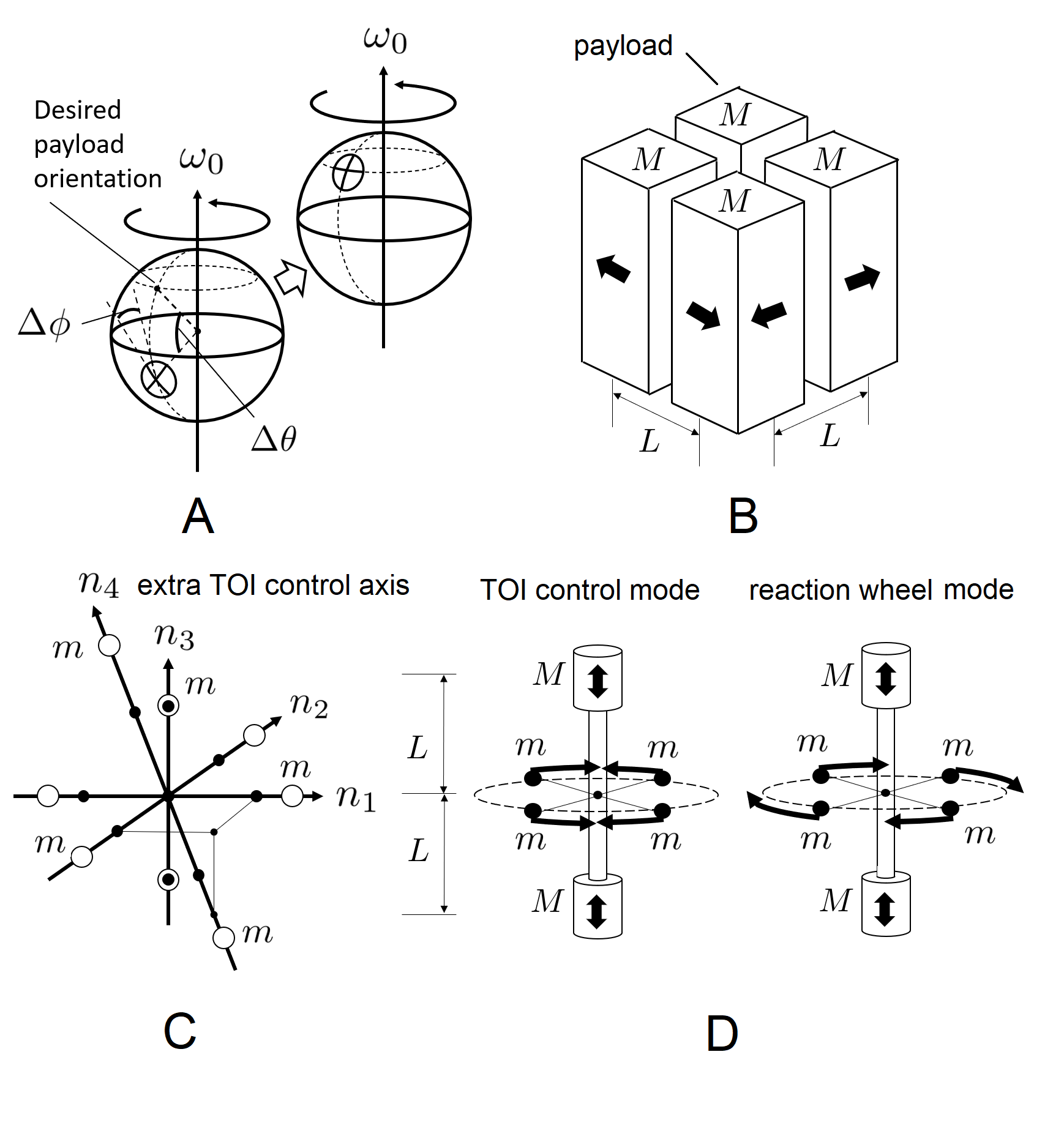}
    \caption{ A) Positioning a payload by spacecraft re-orientation ($\Delta \theta = \theta_{goal} - \theta$, $\Delta \phi = \phi_{goal} - \phi$).} B) Use the payload to control TOI (similar design is discussed in \cite{Trivailo2022}), C) Introduce extra axes of TOI control, D) Design combining functions of TOI control (``Scissors model'' according to \cite{Trivailo2022} ) and reaction wheel.
    \label{fig:6}
\end{figure}

These novel ideas can vary in possible technical embodiment. Fig \ref{fig:6}. (B) demonstrates the possible design considered above, that can be compatible with CubeSat design specifications \cite{CubeSat}. An extremely important feature of the design is that every mass is not a dead weight but a payload - a massive optical objective of an Earth surveillance camera, chemical batteries and other energy storage devices, etc.   

The results above have demonstrated that the controlled development of intermediate axis instability may not be an optimal way to re-orient the spacecraft, especially if the duration of the maneuver should be minimized. An alternative approach is to introduce additional axes of possible TOI control, as illustrated in Fig. \ref{fig:6} (C). Such designs are capable of altering TOI beyond our definition of ``change'' given above, as they allow to instantaneously re-define the directions of principal axes, which enables much faster pre-computed maneuvers to achieve the desirable attitude.

Yet another design illustrated in Fig \ref{fig:6} (D) showcases two important features of the approach. First, the described change in the TOI can be achieved not only via translation of masses, but their counter-rotations. Second, this design illustrates an idea of combination of a functional of a reaction wheel and changing TOI in a single device, depending on whether a co-rotation or counter-rotation of the moving parts is used. Such a design is able to align its axis of rotation of moving parts with the angular velocity by changing TOI, and then stop the rotation of the central shaft by redistributing the angular momentum to the rotating masses acting in a ``reaction wheel'' mode.

It is easy to see that any number of mass translation or counter-rotation mechanisms does not expand the manifold of available stable orientations, and therefore, any re-orientation available with arbitrarily complex mechanism altering TOI can be achieved by the simplest configuration depicted in Fig.1(A) -- although, as mentioned above, more flexible device can perform the required maneuver faster. It is also worth noting that stabilizing rotation of a design depicted in Fig.1(A) with the prescribed orientation $(\theta, \phi)$ that does not match one of mass translation axes, is only possible if the TOI is spherical -- therefore, our approach is the only one providing arbitrary re-orientation for such a class of designs.     

An essential feature of the approach - it can be used in a wide range of angular momenta and self-spin angular velocities - since they only determine the time scale of the maneuver and do not affect its feasibility. Therefore, the same pre-computed maneuver can in principle be performed by a large crewed orbital station and small, rapidly spinning CubeSat.

Our method can be implemented technologically and used in real spacecraft systems. The numerical simulations indicate that the maneuver can always be found, given the sufficient span of control signals $[q_{min}, q_{max}]$, dimensionless duration $T$ and complexity (number of reference control points $N$). The maneuver is resolved with limited precision given insufficient span, duration and/or complexity, and becomes non-unique if the these parameters provided to the optimization procedure are higher than necessary minimum. Except the special case of switching between orientations close to the principal axes, the maneuver can be resolved precisely with reasonable $T$, $N$ and $[q_{min}, q_{max}]$.  

The optimization method showcased in this paper should be viewed only as a proof of concept technique, demonstrating that the maneuvering described above is possible. Using modern machine learning/optimization techniques the method can be substantially improved, producing not only precise, but shortest and the most energy-efficient maneuvers.

As could be seen above, calculation of every maneuver requires considerable computational efforts and can not be done on the fly. Therefore, for practical usage one needs to pre-compute and tabulate all meaningful rotations $(\theta_{beg}, \phi_{beg})\longrightarrow (\theta_{end}, \phi_{end})$. Complete pre-computed table is a 5-d array, and its sufficiently dense sampling and storage poses a problem. To address this challenge, one could take a closer look at structure of the phase space of the rigid motion with the changing TOI, and tabulate this array in adaptive manner. Alternatively, it is possible to use black-box methods, e.g. Tensor Train Cross-approximation \cite{Oseledets2011}, in a manner similar to \cite{Ostanin2017}, to construct a low-rank representation of the complete table of all possible rotations and the corresponding control signals. Given the symmetric structure of this array and its presumable low-rank structure, the black-box approximation should be very efficient tool to accelerate necessary pre-computations. 

One particularly interesting direction is the exploration of small corrections of the attitude. It can be expected that small adjustments (the functional (\ref{eq8}) is initially less than $10^{-1}$) could be achieved by fast maneuvers with very few reference points. The larger maneuvers can then be represented as sequences of smaller ones. 

In this work, the approach was demonstrated only in the numerical simulation. Further study would certainly require manufacturing the demonstration prototypes designed for gyroscope frames, drop towers and parabolic flights, which will pave the way to possible experiments with CubeSats and on-board devices like \cite{Spheres}, toward practical utilization in larger space systems.

The international patent application (Netherlands patent application 2034951, filed 30.05.23) has been submitted prior to the publication. 

All the codes used in this work are available as part of MercuryDPM software \cite{weinhart2020fast, MercuryDPM2023}. They are located in the developer's branch (\burl{https://bitbucket.org/mercurydpm/}) at \texttt{Drivers/Clump/ChangingTOI/}.

\section*{Acknowledgments}

The authors thank Prof. Stefan Luding (Multi-Scale Mechanics (MSM), Faculty of Engineering Technology, MESA+, University of Twente) for the fruitful discussions on the topic.


\newpage
\section*{Supplementary information}
\setcounter{page}{1}
\setcounter{equation}{0}
\setcounter{figure}{0}

\renewcommand{\theequation}{S.\arabic{equation}}

\renewcommand{\thefigure}{S.\arabic{figure}}

\paragraph{Equations of motion of a body that changes its TOI}

The equations of motion of a body changing its TOI are obtained straightforwardly by generalization of standard derivation of Euler's equations of rigid body rotation. In case of zero total external moment acting on the body, time rate-of-change of the angular momentum in the inertial frame of reference should be zero:

\begin{equation} \label{s_eq1}	
	\dot{ \mathbf{L}} = 0  	
\end{equation}

The angular momentum $\mathbf{L}$ is given in the local coordinate frame instantaneously aligned with the body's principal axes as: 

\begin{equation} \label{s_eq2}	
    \mathbf{L}^l = \mathbf{I}^l(t) \boldsymbol{\omega}^l(t)
\end{equation}

In such a local coordinate frame, rotating with the angular velocity $\boldsymbol{\omega}^l(t)$ around its origin, the time derivative of a vector $\mathbf{L}$ is given by:

\begin{equation} \label{s_eq3}	
	(\dot {\mathbf{L}}^l)^{rot} = \dot {\mathbf{L}}^l(t) + \boldsymbol{\omega}^l(t) \times \mathbf{L}^l(t)   
\end{equation}

The equation (\ref{s_eq1}) can therefore be written in a local (rotating) frame as:

\begin{equation} \label{s_eq4}	
	\dot{\mathbf{I}}^l(t) \boldsymbol{\omega}^l(t) + \mathbf{I}^l(t) \dot{\boldsymbol{\omega}}^l(t) + \boldsymbol{\omega}^l(t) \times \mathbf{I}^l(t)\boldsymbol{\omega}^l(t) = 0   
\end{equation}

The equation \ref{s_eq4} can be re-written in inertial Cartesian frame component-wise using indicial notation as: 

\begin{equation} \label{s_eq5}
    \dot I_{ij}(t) \omega_j(t) +
	I_{ij}(t) \dot \omega_j(t) +  \epsilon_{ijk} \omega_j(t) I_{kl}(t) \omega_l(t) = 0 	
\end{equation}
  
The non-spherical TOI $I_{ij}(t)$, its time derivatives $\dot I_{ij}(t)$, the angular velocity $\omega_i(t)$ and its derivative $\dot \omega_i(t)$ are found by proper rotations of the corresponding components in the local frame:

\begin{equation} \label{s_eq6}
	I_{il}(t) = Q_{ik}^T(t) I_{lj}^l(t) Q_{kj}(t),
\end{equation}

\begin{equation} \label{s_eq7}
	\dot{I}_{il}(t) = Q_{ik}^T(t) \dot{I}_{lj}^l(t) Q_{kj}(t),
\end{equation}

\begin{equation} \label{s_eq8}
	\omega_i(t) = Q_{ij}(t) \omega_j^l(t).
\end{equation}

\begin{equation} \label{s_eq9}
	\dot {\omega_i}(t) = Q_{ij}(t) \dot {\omega_j}^l(t).
\end{equation}

The TOI components in local coordinate system $I_{ij}^{l}(t)$ is given by
\begin{equation} \label{s_eq10}	
	\mathbf{I}^{l}(t) = \frac{I_0}{2}\begin{pmatrix}
		1 + q_2(t)^2 & 0 & 0\\
		0 & 1 + q_1(t)^2 & 0\\
		0 & 0 & q_1(t)^2 + q_2(t)^2\\
		\end{pmatrix},
\end{equation}

and $\mathbf{Q}(t)$ is the rotation matrix defined as

\begin{equation} \label{s_eq11}	
	\mathbf{Q}(t) = \begin{pmatrix}
		\mathbf{e}_1\mathbf{n}_1(t) & \mathbf{e}_2\mathbf{n}_1(t) & \mathbf{e}_3\mathbf{n}_1(t)\\
		\mathbf{e}_1\mathbf{n}_2(t) & \mathbf{e}_2\mathbf{n}_2(t) & \mathbf{e}_3\mathbf{n}_2(t)\\
		\mathbf{e}_1\mathbf{n}_3(t) & \mathbf{e}_2\mathbf{n}_3(t) & \mathbf{e}_3\mathbf{n}_3(t)\\
		\end{pmatrix}
\end{equation}

where $\mathbf{e}_i$ are the orths of global Cartesian coordinate system, and $\mathbf{n}_i(t)$ are orths of the body's eigendirections.

\paragraph*{Time integration of the equations of motion}
\medskip


The time integration scheme used in this work utilizes a leap-frog algorithm of the time integration of the motion of non-spherical particle \cite{Ostanin2023}, similar to the algorithm utilized in the commercial software PFC 4.0 \cite{pfc2008}.   
The equation (\ref{s_eq5}) is solved using finite difference procedure of the second order, computing angular velocities $\omega_j$ at mid-intervals $t + \Delta t/2$, and all other dynamic quantities at primary intervals $t + \Delta t$.
The orientation is tracked in the shape of rotation matrix $Q(t)$ that is obtained by the incremental rotation of principal directions by the angle $\boldsymbol{\omega}(t)\Delta t$.
The equation (\ref{s_eq5}) can be re-written in the matrix form as

\begin{equation} \label{s_eq12}
	\begin{split}
		\dot{\mathbf{I}} \boldsymbol{\omega} +
		\mathbf{I} \dot{\boldsymbol{\omega}} +
		\mathbf{W} &= 0,\\
		W &= \begin{pmatrix}
		(I_{33} - I_{22})\omega_2 \omega_3 
		+ I_{23} \omega_3 \omega_3
		- I_{32} \omega_2 \omega_2
		- I_{31} \omega_1 \omega_2
		+ I_{21} \omega_1 \omega_3
		\\
		(I_{11} - I_{33})\omega_3 \omega_1 
		+ I_{31} \omega_1 \omega_1
		- I_{13} \omega_3 \omega_3
		- I_{12} \omega_2 \omega_3
		+ I_{32} \omega_2 \omega_1
		\\
		(I_{22} - I_{11})\omega_1 \omega_2 
		+ I_{12} \omega_2 \omega_2
		- I_{21} \omega_1 \omega_1
		- I_{23} \omega_3 \omega_1
		+ I_{13} \omega_3 \omega_2
		\\
		\end{pmatrix}, \\
		I &= \begin{pmatrix}
		 I_{11} & -I_{12} & -I_{13}\\
		-I_{21} &  I_{22} & -I_{23}\\
		-I_{31} & -I_{32} &  I_{33}\\
		\end{pmatrix}, \\
	    \dot{I} &= \begin{pmatrix}
		 \dot{I}_{11} & -\dot{I}_{12} & -\dot{I}_{13}\\
		-\dot{I}_{21} &  \dot{I}_{22} & -\dot{I}_{23}\\
		-\dot{I}_{31} & -\dot{I}_{32} &  \dot{I}_{33}\\
		\end{pmatrix}.
	\end{split}
\end{equation}

Here the regular sign convention for tensor of inertia components is implied. Using three equations (\ref{s_eq12}), we need to evaluate six unknowns $\omega_i(t+\Delta t/2)$, $\dot \omega_i(t+\Delta t)$. Following the approach suggested in \cite{pfc2008} we use the iterative algorithm to find these unknowns:

\begin{itemize}
    \item Set $n=0$ \\
    \item Set $\omega_i^{[0]}$ to the initial angular velocity.
    \item (*) Solve (\ref{s_eq12}) for $\dot \omega_i$
    \item Determine a new (intermediate) angular velocity: $\omega_i^{[new]} = \omega_i^{[0]} + \dot \omega_i^{[n]} \Delta t$
    \item Revise the estimate of $\omega_i$ as:
    $\omega_i^{[n+1]} = 0.5 (\omega_i^{[0]} + \omega_i^{[new]})$
    \item Set $n:=n+1$ and go to (*)
\end{itemize}

This algorithm gives us the value of the angular velocity that is further used to update the position at the second step of leap-frog algorithm. The number of steps $n$ necessary for the sufficient precision varies depending on the application and, based on our numerical experiments, in our simulations it was set to $3$.

The timestep in our simulations have been chosen rather fine - approximately $450$ timesteps per single revolution of the body, given $q_1 = q_2 = 1$. Such a timestep was chosen empirically to ensure negligible dependence of the simulation-guided optimization results on the timestep.  

The described algorithm is rather simple, however, the numerical results demonstrate that in all the maneuvers described in this work it conserves the magnitude and direction of angular momentum nearly precisely, and the relative drift of the energy between same-energy initial and final state appears to be vanishingly small - see Section \ref{Results}.

\paragraph{Supplementary videos}

\begin{figure} 
    \centering
    \includegraphics[width=\textwidth]{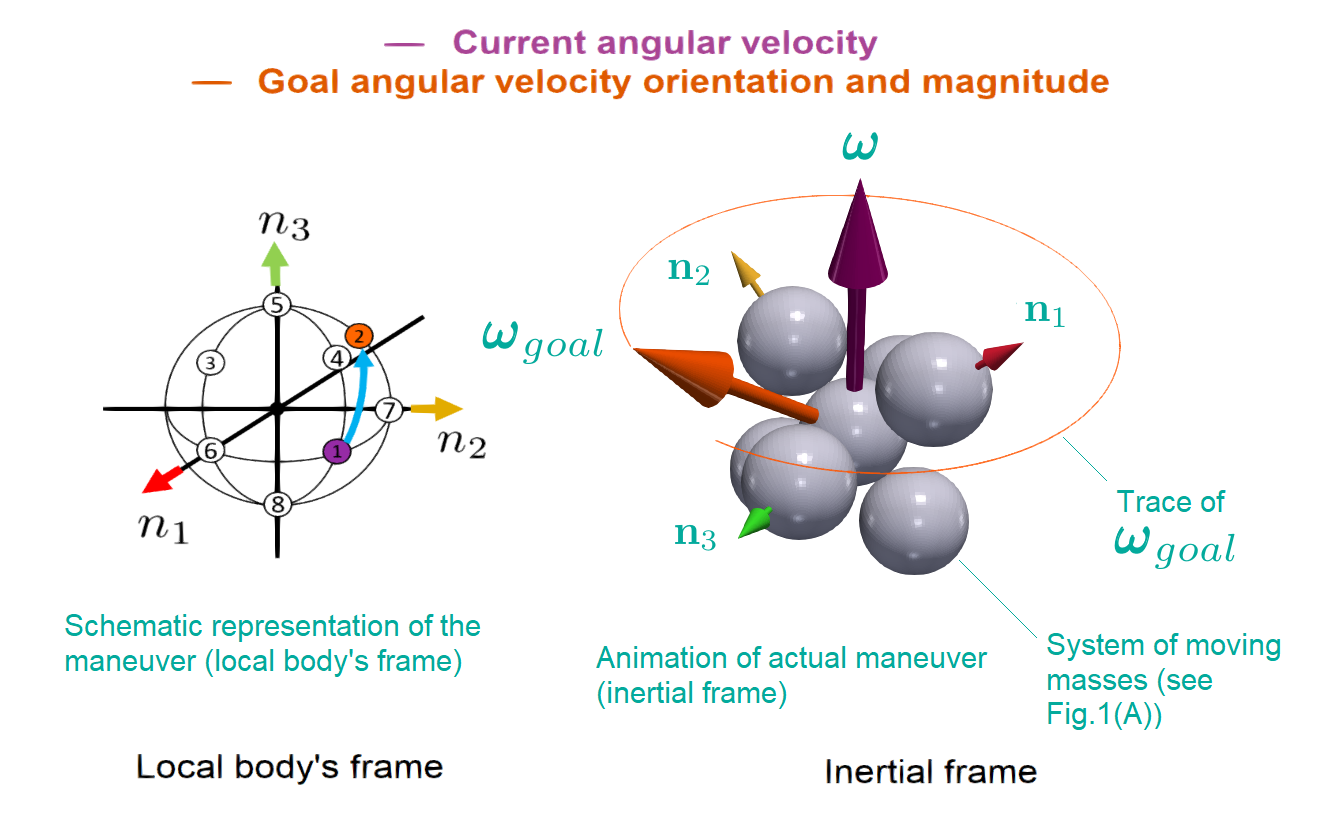} 
    \caption{Explanation of supplementary videos layout. Annotations are given in green).}
    \label{fig:1s}
\end{figure}

The videos demonstrate time evolution of the body's orientation during the maneuver. Fig. \ref{fig:1s} gives the video layout with the annotations (highlighted with green). Each video showcases the scheme of a featured maneuver, depicted as the rotation of the direction of the angular velocity in local coordinate system (on the right side of the layout). The animation gives the body, represented in inertial coordinate system with three couples of spherical particles (sketching the mechanism similar to Fig. \ref{fig:1}(A)) plus the particle in the center of mass. The drawing also includes local axes $n_1,n_2,n_3$ indicated with red, yellow and green vectors, correspondingly. The orange vector (constant in the body's coordinates) represents the desired orientation and magnitude of the angular velocity in the body's local coordinate system:

\begin{equation} \label{s_eq13}	
	\mathbf{\omega}_{goal} = \left| \omega_0 \right| \mathbf{p}_{goal}(\theta_{goal}, \phi_{goal})
\end{equation}

The time evolution of $\boldsymbol{\omega}_{goal}$ is traced with the orange streamline. The purple vector denotes the current angular velocity $\omega_0$. At the end of the successful maneuver $\boldsymbol{\omega}_{goal} = \boldsymbol{\omega}_{0}$. The videos are offered for few maneuvers showcased in Table 1.  

\begin{thebibliography}{29}
	\ifx \bisbn   \undefined \def \bisbn  #1{ISBN #1}\fi
	\ifx \binits  \undefined \def \binits#1{#1}\fi
	\ifx \bauthor  \undefined \def \bauthor#1{#1}\fi
	\ifx \batitle  \undefined \def \batitle#1{#1}\fi
	\ifx \bjtitle  \undefined \def \bjtitle#1{#1}\fi
	\ifx \bvolume  \undefined \def \bvolume#1{\textbf{#1}}\fi
	\ifx \byear  \undefined \def \byear#1{#1}\fi
	\ifx \bissue  \undefined \def \bissue#1{#1}\fi
	\ifx \bfpage  \undefined \def \bfpage#1{#1}\fi
	\ifx \blpage  \undefined \def \blpage #1{#1}\fi
	\ifx \burl  \undefined \def \burl#1{\textsf{#1}}\fi
	\ifx \doiurl  \undefined \def \doiurl#1{\url{https://doi.org/#1}}\fi
	\ifx \betal  \undefined \def \betal{\textit{et al.}}\fi
	\ifx \binstitute  \undefined \def \binstitute#1{#1}\fi
	\ifx \binstitutionaled  \undefined \def \binstitutionaled#1{#1}\fi
	\ifx \bctitle  \undefined \def \bctitle#1{#1}\fi
	\ifx \beditor  \undefined \def \beditor#1{#1}\fi
	\ifx \bpublisher  \undefined \def \bpublisher#1{#1}\fi
	\ifx \bbtitle  \undefined \def \bbtitle#1{#1}\fi
	\ifx \bedition  \undefined \def \bedition#1{#1}\fi
	\ifx \bseriesno  \undefined \def \bseriesno#1{#1}\fi
	\ifx \blocation  \undefined \def \blocation#1{#1}\fi
	\ifx \bsertitle  \undefined \def \bsertitle#1{#1}\fi
	\ifx \bsnm \undefined \def \bsnm#1{#1}\fi
	\ifx \bsuffix \undefined \def \bsuffix#1{#1}\fi
	\ifx \bparticle \undefined \def \bparticle#1{#1}\fi
	\ifx \barticle \undefined \def \barticle#1{#1}\fi
	\bibcommenthead
	\ifx \bconfdate \undefined \def \bconfdate #1{#1}\fi
	\ifx \botherref \undefined \def \botherref #1{#1}\fi
	\ifx \url \undefined \def \url#1{\textsf{#1}}\fi
	\ifx \bchapter \undefined \def \bchapter#1{#1}\fi
	\ifx \bbook \undefined \def \bbook#1{#1}\fi
	\ifx \bcomment \undefined \def \bcomment#1{#1}\fi
	\ifx \oauthor \undefined \def \oauthor#1{#1}\fi
	\ifx \citeauthoryear \undefined \def \citeauthoryear#1{#1}\fi
	\ifx \endbibitem  \undefined \def \endbibitem {}\fi
	\ifx \bconflocation  \undefined \def \bconflocation#1{#1}\fi
	\ifx \arxivurl  \undefined \def \arxivurl#1{\textsf{#1}}\fi
	\csname PreBibitemsHook\endcsname
	
	\bibitem[\protect\citeauthoryear{Trivailo and Kojima}{2022}]{Trivailo2022}
	\begin{bbook}
		\bauthor{\bsnm{Trivailo}, \binits{P.M.}},
		\bauthor{\bsnm{Kojima}, \binits{H.}}:
		In: \beditor{\bsnm{Dai}, \binits{L.}},
		\beditor{\bsnm{Jazar}, \binits{R.N.}} (eds.)
		\bbtitle{Inertial Morphing as a Novel Concept in Attitude Control and Design of Variable Agility Acrobatic Autonomous Spacecraft},
		pp. \bfpage{119}--\blpage{244}.
		\bpublisher{Springer},
		\blocation{Cham}
		(\byear{2022}).
		\doiurl{10.1007/978-3-030-82719-9_5} .
		\burl{https://doi.org/10.1007/978-3-030-82719-9_5}
	\end{bbook}
	\endbibitem
	
	\bibitem[\protect\citeauthoryear{Poinsot}{1834}]{Poinsot1834}
	\begin{bbook}
		\bauthor{\bsnm{Poinsot}, \binits{L.}}:
		\bbtitle{Theorie Nouvelle de la Rotation des Corps}.
		\bpublisher{Bachelier},
		\blocation{Paris}
		(\byear{1834})
	\end{bbook}
	\endbibitem
	
	\bibitem[\protect\citeauthoryear{Wertz}{1978}]{Wertz_1978}
	\begin{bbook}
		\bauthor{\bsnm{Wertz}, \binits{J.R.}}:
		\bbtitle{Spacecraft Attitude Determination and Control},
		\bedition{7}th edn.
		\bpublisher{Springer},
		\blocation{Dordrecht}
		(\byear{1978})
	\end{bbook}
	\endbibitem
	
	\bibitem[\protect\citeauthoryear{Hughes}{2004}]{Hughes_2004}
	\begin{bbook}
		\bauthor{\bsnm{Hughes}, \binits{P.C.}}:
		\bbtitle{Spacecraft Attitude Dynamics}.
		\bpublisher{Dover Publications},
		\blocation{New York}
		(\byear{2004})
	\end{bbook}
	\endbibitem
	
	\bibitem[\protect\citeauthoryear{{Van Damme} et~al.}{2017}]{VanDamme2017}
	\begin{barticle}
		\bauthor{\bsnm{{Van Damme}}, \binits{L.}},
		\bauthor{\bsnm{Mardešić}, \binits{P.}},
		\bauthor{\bsnm{Sugny}, \binits{D.}}:
		\batitle{The tennis racket effect in a three-dimensional rigid body}.
		\bjtitle{Physica D: Nonlinear Phenomena}
		\bvolume{338},
		\bfpage{17}--\blpage{25}
		(\byear{2017})
		\doiurl{10.1016/j.physd.2016.07.010}
	\end{barticle}
	\endbibitem
	
	\bibitem[\protect\citeauthoryear{Beachley and Uicker}{1969}]{Beachley1969}
	\begin{barticle}
		\bauthor{\bsnm{Beachley}, \binits{N.H.}},
		\bauthor{\bsnm{Uicker}, \binits{J.J.}}:
		\batitle{Reply by authors to l. h. grasshoff}.
		\bjtitle{Journal of Spacecraft and Rockets}
		\bvolume{6}(\bissue{10}),
		\bfpage{1215}--\blpage{1216}
		(\byear{1969})
		\doiurl{10.2514/3.59631}
		{\href{https://arxiv.org/abs/https://doi.org/10.2514/3.59631}{{https://doi.org/10.2514/3.59631}}}
	\end{barticle}
	\endbibitem
	
	\bibitem[\protect\citeauthoryear{Beachley}{1971}]{Beachley1971}
	\begin{barticle}
		\bauthor{\bsnm{Beachley}, \binits{N.H.}}:
		\batitle{Inversion of spin-stabilized spacecraft by mass translation - some practical aspects}.
		\bjtitle{Journal of Spacecraft and Rockets}
		\bvolume{8}(\bissue{10}),
		\bfpage{1078}--\blpage{1080}
		(\byear{1971})
		\doiurl{10.2514/3.30349}
		{\href{https://arxiv.org/abs/https://doi.org/10.2514/3.30349}{{https://doi.org/10.2514/3.30349}}}
	\end{barticle}
	\endbibitem
	
	\bibitem[\protect\citeauthoryear{Kane and Sobala}{1963}]{Kane1963}
	\begin{barticle}
		\bauthor{\bsnm{Kane}, \binits{T.R.}},
		\bauthor{\bsnm{Sobala}, \binits{D.}}:
		\batitle{A new method for attitude stabilization}.
		\bjtitle{AIAA Journal}
		\bvolume{1}(\bissue{6}),
		\bfpage{1365}--\blpage{1367}
		(\byear{1963})
		\doiurl{10.2514/3.1794}
		{\href{https://arxiv.org/abs/https://doi.org/10.2514/3.1794}{{https://doi.org/10.2514/3.1794}}}
	\end{barticle}
	\endbibitem
	
	\bibitem[\protect\citeauthoryear{Edwards and Kaplan}{1974}]{Edwards1974}
	\begin{barticle}
		\bauthor{\bsnm{Edwards}, \binits{T.L.}},
		\bauthor{\bsnm{Kaplan}, \binits{M.H.}}:
		\batitle{Automatic spacecraft detumbling by internal mass motion}.
		\bjtitle{AIAA Journal}
		\bvolume{12}(\bissue{4}),
		\bfpage{496}--\blpage{502}
		(\byear{1974})
		\doiurl{10.2514/3.49275}
		{\href{https://arxiv.org/abs/https://doi.org/10.2514/3.49275}{{https://doi.org/10.2514/3.49275}}}
	\end{barticle}
	\endbibitem
	
	\bibitem[\protect\citeauthoryear{Mayorova et~al.}{2011}]{Mayorova2011patent}
	\begin{botherref}
		\oauthor{\bsnm{Mayorova}, \binits{V.I.}},
		\oauthor{\bsnm{Popov}, \binits{A.S.}},
		\oauthor{\bsnm{Tenenbaum}, \binits{S.M.}},
		\oauthor{\bsnm{Kotsur}, \binits{O.S.}},
		\oauthor{\bsnm{Rachkin}, \binits{D.A.}},
		\oauthor{\bsnm{Nerovny}, \binits{N.A.}},
		\oauthor{\bsnm{Nazarov}, \binits{N.G.}}:
		Method for reorientating and controlling the thrust of a rotating spacecraft with a solar sail.
		Google Patents.
		WIPO WO2013002673A1
		(2011).
		\url{https://patents.google.com/patent/WO2013002673A1/en}
	\end{botherref}
	\endbibitem
	
	\bibitem[\protect\citeauthoryear{Trivailo}{2017}]{Trivailo2017}
	\begin{bchapter}
		\bauthor{\bsnm{Trivailo}, \binits{P.M.}}:
		\bctitle{Utilisation of the “dzhanibekov’s effect” for the possible future space missions}.
		In: \bbtitle{Proc. of 26-th Int. Symp. Sp. Fl. Dyn., Matsuyama, Japan},
		pp. \bfpage{1}--\blpage{10}
		(\byear{2017})
	\end{bchapter}
	\endbibitem
	
	\bibitem[\protect\citeauthoryear{Weinhart et~al.}{2020}]{weinhart2020fast}
	\begin{barticle}
		\bauthor{\bsnm{Weinhart}, \binits{T.}},
		\bauthor{\bsnm{Orefice}, \binits{L.}},
		\bauthor{\bsnm{Post}, \binits{M.}},
		\bauthor{\bsnm{Schrojenstein~Lantman}, \binits{M.P.}},
		\bauthor{\bsnm{Denissen}, \binits{I.}},
		\bauthor{\bsnm{Tunuguntla}, \binits{D.R.}},
		\bauthor{\bsnm{Tsang}, \binits{J.M.F.}},
		\bauthor{\bsnm{Cheng}, \binits{H.}},
		\bauthor{\bsnm{Shaheen}, \binits{M.Y.}},
		\bauthor{\bsnm{Shi}, \binits{H.}}:
		\batitle{Fast, flexible particle simulations -- an introduction to {MercuryDPM}}.
		\bjtitle{Computer Physics Communications}
		\bvolume{249},
		\bfpage{107129}
		(\byear{2020})
	\end{barticle}
	\endbibitem
	
	\bibitem[\protect\citeauthoryear{Richter}{2006}]{Richter2006}
	\begin{barticle}
		\bauthor{\bsnm{Richter}, \binits{P.H.}}:
		\batitle{Regular and chaotic rigid body dynamics}.
		\bjtitle{Nonlinear Phenomena in Complex Systems}
		\bvolume{9}(\bissue{2}),
		\bfpage{115}--\blpage{124}
		(\byear{2006})
	\end{barticle}
	\endbibitem
	
	\bibitem[\protect\citeauthoryear{Ashbaugh et~al.}{1991}]{Ashbaugh1991}
	\begin{barticle}
		\bauthor{\bsnm{Ashbaugh}, \binits{M.S.}},
		\bauthor{\bsnm{Chicone}, \binits{C.C.}},
		\bauthor{\bsnm{Cushman}, \binits{R.H.}}:
		\batitle{The twisting tennis racket}.
		\bjtitle{Journal of Dynamics and Differential Equations}
		\bvolume{3}(\bissue{1}),
		\bfpage{67}--\blpage{85}
		(\byear{1991})
		\doiurl{10.1007/BF01049489}
	\end{barticle}
	\endbibitem
	
	\bibitem[\protect\citeauthoryear{Ostanin et~al.}{2022}]{Ostanin2022}
	\begin{barticle}
		\bauthor{\bsnm{Ostanin}, \binits{I.}},
		\bauthor{\bsnm{Cheng}, \binits{H.}},
		\bauthor{\bsnm{Magnanimo}, \binits{V.}}:
		\batitle{Simulation-guided optimization of granular phononic crystal structure using the discrete element method}.
		\bjtitle{Extreme Mechanics Letters}
		\bvolume{55},
		\bfpage{101825}
		(\byear{2022})
	\end{barticle}
	\endbibitem
	
	\bibitem[\protect\citeauthoryear{{Virtanen} and {Contributors}}{2020}]{2020SciPy}
	\begin{barticle}
		\bauthor{\bsnm{{Virtanen}}, \binits{P.}},
		\bauthor{\bsnm{{Contributors}}, \binits{S...}}:
		\batitle{{SciPy 1.0: Fundamental Algorithms for Scientific Computing in Python}}.
		\bjtitle{Nature Methods}
		\bvolume{17},
		\bfpage{261}--\blpage{272}
		(\byear{2020})
		\doiurl{10.1038/s41592-019-0686-2}
	\end{barticle}
	\endbibitem
	
	\bibitem[\protect\citeauthoryear{Powell}{1964}]{Powell1964}
	\begin{barticle}
		\bauthor{\bsnm{Powell}, \binits{M.J.D.}}:
		\batitle{An efficient method for finding the minimum of a function of several variables without calculating derivatives}.
		\bjtitle{Computer Journal.}
		\bvolume{7},
		\bfpage{155}--\blpage{162}
		(\byear{1964})
		\doiurl{10.1093/comjnl/7.2.155}
	\end{barticle}
	\endbibitem
	
	\bibitem[\protect\citeauthoryear{}{2023a}]{Maneuver_1}
	\begin{botherref}
		Supplementary Information. Video of maneuver 1.
		\url{https://youtu.be/GNt-SibjcgE}
		(2023)
	\end{botherref}
	\endbibitem
	
	\bibitem[\protect\citeauthoryear{}{2023b}]{Maneuver_2}
	\begin{botherref}
		Supplementary Information. Video of maneuver 2.
		\url{https://youtu.be/9_fMZeLiooo}
		(2023)
	\end{botherref}
	\endbibitem
	
	\bibitem[\protect\citeauthoryear{}{2023c}]{Maneuver_3}
	\begin{botherref}
		Supplementary Information. Video of maneuver 3.
		\url{https://youtu.be/Gj6wN_eWf90}
		(2023)
	\end{botherref}
	\endbibitem
	
	\bibitem[\protect\citeauthoryear{}{2023d}]{Maneuver_4}
	\begin{botherref}
		Supplementary Information. Video of maneuver 4.
		\url{https://youtu.be/gyiRDuj9wgI}
		(2023)
	\end{botherref}
	\endbibitem
	
	\bibitem[\protect\citeauthoryear{}{2023e}]{Maneuver_10}
	\begin{botherref}
		Supplementary Information. Video of maneuver 10.
		\url{https://youtu.be/yfrtPekKFzA}
		(2023)
	\end{botherref}
	\endbibitem
	
	\bibitem[\protect\citeauthoryear{}{2023}]{CubeSat}
	\begin{botherref}
		CubeSat design specification.
		\url{https://static1.squarespace.com/static/5418c831e4b0fa4ecac1bacd/t/62193b7fc9e72e0053f00910/1645820809779/CDS+REV14_1+2022-02-09.pdf}
		(2023)
	\end{botherref}
	\endbibitem
	
	\bibitem[\protect\citeauthoryear{Oseledets}{2011}]{Oseledets2011}
	\begin{barticle}
		\bauthor{\bsnm{Oseledets}, \binits{I.V.}}:
		\batitle{Tensor-train decomposition}.
		\bjtitle{SIAM Journal on Scientific Computing}
		\bvolume{33}(\bissue{5}),
		\bfpage{2295}--\blpage{2317}
		(\byear{2011})
		\doiurl{10.1137/090752286}
		{\href{https://arxiv.org/abs/https://doi.org/10.1137/090752286}{{https://doi.org/10.1137/090752286}}}
	\end{barticle}
	\endbibitem
	
	\bibitem[\protect\citeauthoryear{Drozdov et~al.}{2017}]{Ostanin2017}
	\begin{barticle}
		\bauthor{\bsnm{Drozdov}, \binits{G.}},
		\bauthor{\bsnm{Ostanin}, \binits{I.}},
		\bauthor{\bsnm{Oseledets}, \binits{I.}}:
		\batitle{Time-and memory-efficient representation of complex mesoscale potentials}.
		\bjtitle{Journal of Computational Physics}
		\bvolume{343},
		\bfpage{110}--\blpage{114}
		(\byear{2017})
	\end{barticle}
	\endbibitem
	
	\bibitem[\protect\citeauthoryear{}{2023}]{Spheres}
	\begin{botherref}
		SPHERES mission.
		https://www.nasa.gov/spheres/home
		(2023)
	\end{botherref}
	\endbibitem
	
	\bibitem[\protect\citeauthoryear{Thornton et~al.}{2023}]{MercuryDPM2023}
	\begin{barticle}
		\bauthor{\bsnm{Thornton}, \binits{A.R.}},
		\bauthor{\bsnm{Plath}, \binits{T.}},
		\bauthor{\bsnm{Ostanin}, \binits{I.}},
		\bauthor{\bsnm{G{\"o}tz}, \binits{H.}},
		\bauthor{\bsnm{Bisschop}, \binits{J.-W.}},
		\bauthor{\bsnm{Hassan}, \binits{M.}},
		\bauthor{\bsnm{Roeplal}, \binits{R.}},
		\bauthor{\bsnm{Wang}, \binits{X.}},
		\bauthor{\bsnm{Pourandi}, \binits{S.}},
		\bauthor{\bsnm{Weinhart}, \binits{T.}}:
		\batitle{Recent advances in mercurydpm}.
		\bjtitle{Mathematics in Computer Science}
		\bvolume{17}(\bissue{2}),
		\bfpage{13}
		(\byear{2023})
		\doiurl{10.1007/s11786-023-00562-x}
	\end{barticle}
	\endbibitem
	
	\bibitem[\protect\citeauthoryear{Ostanin et~al.}{2023}]{Ostanin2023}
	\begin{botherref}
		\oauthor{\bsnm{Ostanin}, \binits{I.}},
		\oauthor{\bsnm{Angelidakis}, \binits{V.}},
		\oauthor{\bsnm{Plath}, \binits{T.}},
		\oauthor{\bsnm{Pourandi}, \binits{S.}},
		\oauthor{\bsnm{Thornton}, \binits{A.}},
		\oauthor{\bsnm{Weinhart}, \binits{T.}}:
		Rigid clumps in the mercurydpm particle dynamics code
		(2023)
		{\href{https://arxiv.org/abs/2310.05027}{{arXiv:2310.05027}}}
	\end{botherref}
	\endbibitem
	
	\bibitem[\protect\citeauthoryear{Inc.}{2008}]{pfc2008}
	\begin{botherref}
		\oauthor{\bsnm{Inc.}, \binits{I.C.G.}}:
		PFC3D (Particle Flow Code in 3 Dimensions). Version 4.0.
		Itasca Consulting Group Inc., Minneapolis.
		(2008)
	\end{botherref}
	\endbibitem
	
\end{thebibliography}
\end{document}